\documentclass[12pt,a4paper]{article}
\usepackage{epsfig}
\usepackage{psfrag}

\title{Phase diagram of the mean field model of simplicial  gravity}
\author{P.~Bialas${}^a$\thanks{Permanent address~: Institute of Comp.~Science
    Jagellonian University, 30-072 Krakow, Poland}, Z.~Burda${}^b$ and D.~Johnston${}^c$\\\\
\small ${}^a$Fakult\"at fur Physik, Universit\"at Bielefeld, 33501 Bielefeld Germany\\
\small ${}^b$Institute of Physics, Jagellonian University, 30-059 Krakow, Poland\\
\small ${}^c$ Dept. of Mathematics, Heriot--Watt University, EH14 4AS Edinburgh, Scotland}

\date{}
\begin{document}

\maketitle
\begin{abstract}
  We discuss the phase diagram of the balls in boxes model, with a varying
  number of boxes. The model can be regarded as a mean-field model of simplicial
  gravity.  We analyse in detail the case of weights of the form $p(q) =
  q^{-\beta}$, which correspond to the measure term introduced in the simplicial
  quantum gravity simulations.  The system has two phases~: {\em elongated}
  ({\em fluid}) and {\em crumpled}.  For $\beta\in (2,\infty)$ the transition
  between these two phases is first order, while for $\beta \in (1,2]$ it is
  continuous.  The transition becomes softer when $\beta$ approaches unity and
  eventually disappears at $\beta=1$. We then generalise the discussion to an
  arbitrary set of weights. Finally, we show that if one introduces an
  additional kinematic bound on the average density of balls per box then a new
  {\em condensed} phase appears in the phase diagram.  It bears some similarity
  to the {\em crinkled} phase of simplicial gravity discussed recently in models
  of gravity interacting with matter fields.
\end{abstract}

\section*{}

The mean field description of simplicial gravity introduced in
\cite{bbpt,bbj,bb} was analysed in the series of papers \cite{ckr1,ckr2,zb}.
As discussed there, the mean field approximation explains many facts observed
in numerical experiments of simplicial gravity, such as 
the appearance of singular
vertices and the mother universe, and the discontinuity of the phase transition.
Recent work \cite{bbkptt1,bbkptt2} shows that adding vector matter fields
to simplicial gravity 
effectively amounts to renormalising the $\alpha$ parameter 
in the measure term $q^\alpha$ where $q$ is the vertex order.
The resulting phase diagram resembles the phase diagram of the mean--field
model which  we discuss in detail in this letter.

The model is given by the partition function~\cite{bb}~:
\begin{eqnarray}
Z(N,\kappa,\beta) &=& \sum_{M=1}^{M_{max}} e^{\kappa M}
\sum_{q_1,\ldots,q_M} p(q_1)\cdots p(q_M)\delta{q_1+\cdots+q_M,N} \\
&=& \sum_{M=1}^{M_{max}} e^{ \kappa M} z(N,M,\beta) 
\label{z}
\end{eqnarray}
It describes an ensemble of $N$ balls distributed in
a varying  number of boxes, $M$. The system has at least one box and 
at most $M_{max}$ boxes.  
The partitions of balls are weighted by the product of one-box weights 
$p(q)$, where $q$ is the number of balls in the box. Here
we consider one parameter family of weights~:
\begin{eqnarray}
p(q) = q^{-\beta}\quad\mbox{for}\quad q=1,2,\dots \, .
\label{w}
\end{eqnarray}
Note that for these weights the minimal number 
of balls in a box is $q_{min}=1$. If there are no further
constraints this implies that the maximal number of boxes
is equal to the number of balls $M_{max}=N$. 

For large $N$ the partition functions $z(\ldots)$ and $Z(\ldots)$ 
are expected to behave as~:
\begin{eqnarray}
z(N,M,\beta)  \sim N^y e^{N f(r,\beta) } \quad\mbox{with} \quad  r=\frac{M}{N}\,,
\label{zf}
\end{eqnarray}
and 
\begin{eqnarray}
Z(N,\kappa,\beta)  \sim N^Y e^{N \phi(\kappa,\beta)} 
\label{phi}
\end{eqnarray}
where $f(r,\beta)$ and $\phi(\kappa,\beta)$ are
appropriate thermodynamic potential densities 
depending on the intensive quantities characteristic
for the $(r,\beta)$ and $(\kappa,\beta)$ ensembles. 
The power law corrections for the partition functions
with the exponents $y$, $Y$
disappear in the thermodynamic
limit. In our case they are the leading
corrections. In general, however, there might be stronger corrections,
for example of the type $\exp c N^{\alpha} $, with $\alpha<1$.

Substituting the summation in (\ref{z}) by an integration
over a continuous variable,
we can relate the partition functions by the transform~:
\begin{eqnarray}
N^Y e^{ N \phi(\kappa,\beta)} = 
N^{y+1} \, \int\limits_{0}^{1}\!\!\mbox{d}r e^{N (f(r,\beta) + r\kappa)} \, ,
\label{integ}
\end{eqnarray}
The additional factor $N$ on the right hand side of (\ref{integ})
comes from the integration measure $N dr$.  
In the thermodynamic limit, this gives~:
\begin{eqnarray}
\phi(\kappa,\beta) = f(r_*,\beta) + r_*\kappa \, .
\end{eqnarray}
where $r_*$ is maximum of the integrand in (\ref{integ}).
The value of $r_*$ is given by the equation~:
\begin{eqnarray}
- \kappa = \partial_r f(r_*,\beta)
\label{trans}
\end{eqnarray}
if the maximum of the integrand lies inside
the interval $(0,1)$ and is equal $r_*=0$ or $r_*=1$ 
otherwise. Additionally we see that in the case when the 
maximum lies inside the interval $(0,1)$ the integration
over $r$ around the saddle point introduces the 
additional factor $N^{-1/2}$. This leads
to the following relation between the exponents $y$ and $Y$~:
\begin{eqnarray}
Y = y+\frac{1}{2} \,.
\label{Yy1}
\end{eqnarray}
Otherwise, the maximum lies at either end of the interval 
and the integration over $r$ introduces a factor $N^{-1}$, 
thus
\begin{eqnarray}
Y = y \,.
\label{Yy2}
\end{eqnarray}
We shall return to the exponents $Y,y$ later, concentrating now 
on the thermodynamic limit.

In the $(\kappa,\beta)$ ensemble one defines 
the average $\langle r \rangle$~:
\begin{eqnarray}
\langle r \rangle  = \frac{\langle M \rangle}{N} = 
\frac{1}{N} 
\frac{\partial_\kappa Z(N,\kappa,\beta)}{Z(N,\kappa,\beta)} \, .
\label{ra}
\end{eqnarray}
As we will see this quantity plays the role of the order parameter.
In the thermodynamic limit the equation (\ref{ra}) reads~:
\begin{eqnarray}
\langle r \rangle = \partial_\kappa \phi(\kappa,\beta) = r_*  \, .
\label{itrans}
\end{eqnarray}
The second equality in the last formula can be treated as an inverse transform 
to (\ref{trans}). It results from the fact
that the integrand of (\ref{integ}), which corresponds to the
distribution of $r$, is sharply peaked around 
$r_*$ when $N\rightarrow\infty$ and the position
of the peak also becomes the average of the distribution.

The task of determining $r_*$ relies on finding the 
maximum of the function $f(r,\beta) + \kappa r$. 
The function $f(r,\beta)$ can be found by the saddle
point method \cite{bbj}~:
\begin{eqnarray}
f(r,\beta) = \mu_*(r) + r K(\mu_*(r),\beta)
\label{ff}
\end{eqnarray}
where $\mu_*(r)$ is a value which 
maximises the right hand side of the above expression
for the given $r$.
The function
$K(\mu,\beta)$ is defined as~:
\begin{eqnarray}
K(\mu,\beta) = \log \sum_{q=1}^{\infty} p(q) e^{-\mu q}\, . 
\label{kg0}
\end{eqnarray}

The properties of the function $K(\mu,\beta)$ are central
to the further considerations as they fully determine the behaviour 
of the model. 
For the particular choice 
of weights (\ref{w})~:
\begin{eqnarray}\label{kg}
K(\mu,\beta) = \log \sum_{q=1}^{\infty} q^{-\beta} e^{-\mu q} 
= \log g_{\beta} (\mu)
\end{eqnarray}
where $g_{\beta}(\mu)$ is a function familiar from
Bose-Einstein condensation, which has an
integral representation~:
\begin{eqnarray}
g_\beta(\mu) = \frac{1}{\Gamma(\beta)} \int_0^\infty 
\frac{t^{\beta-1} d t}{e^{\mu+t} - 1} \, .
\label{gb}
\end{eqnarray}
The function $g_\beta(\mu)$ is defined on the interval $[0,\infty)$,
so $\mu_*$ is either a solution of the saddle point equation~:
\begin{eqnarray}\label{mur}
\frac{1}{r} = -{\partial_\mu K(\mu_*,\beta)} = 
\frac{g_{\beta-1}(\mu_*)}{g_{\beta}(\mu_*)}
\end{eqnarray}
for $r > r_{cr}$ or else $\mu_*=0$. The critical value of $r$ is
\begin{eqnarray}
r_{cr}=\left\{
\begin{array}{lr}
0& \beta\le 2\\
\label{rcr}
\frac{\zeta(\beta)}{\zeta(\beta-1)}&\beta>2
\end{array}\right. \, ,
\end{eqnarray} 
where $\zeta(\beta)$ is the Riemann Zeta function which 
arises 
when one inserts $\mu_*=0$ into $K(\mu,\beta)$ in (\ref{kg}).
For $r < r_{cr}$ the function $\mu + r K(\mu,\beta)$ has no maximum in
the range $(0,\infty)$ and $\mu_*=0$.

\begin{figure}[t]
\begin{center}  
\epsfig{file=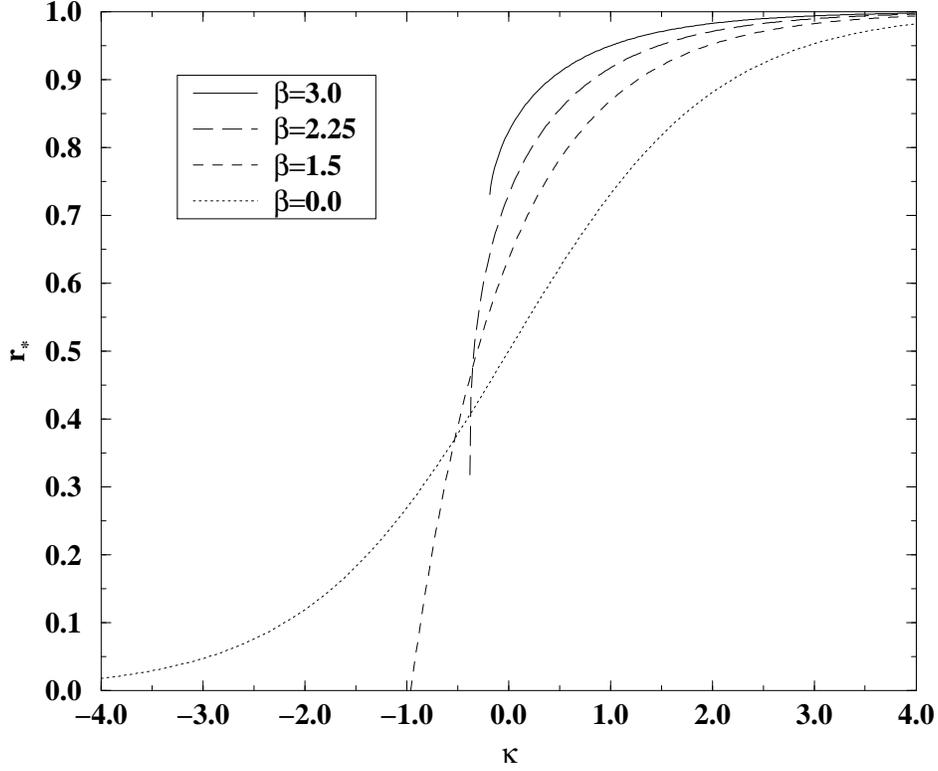,width=12cm,bbllx=51,bblly=60,bburx=516,bbury=434}
\end{center}
\caption{\label{f2}The value of $r_*(\kappa)$ for various values of $\beta$. 
The portions of plots lying on the $x$ axis ($r_*=0$) are not plotted.}
\end{figure}

We thus see that the function $f(r,\beta)$ as a
function of $r$ changes regime at $r=r_{cr}$.
For $r < r_{cr}$ {\em ie} $\mu_*=0$ the equation (\ref{ff})
reduces to the linear dependence~:
\begin{eqnarray}
f(r,\beta) = -\kappa_{cr}(\beta) \, r 
\end{eqnarray}
where
\begin{eqnarray}
-\kappa_{cr} = \log \zeta(\beta)  \, .
\label{kcr}
\end{eqnarray}
We have added the minus sign in the definition of $\kappa_{cr}$  
for convenience to adjust to the sign 
in the saddle point formula (\ref{trans}).
\begin{figure}[t]
\begin{center}
\psfrag{df}{$f(r,\beta)$}  
\epsfig{file=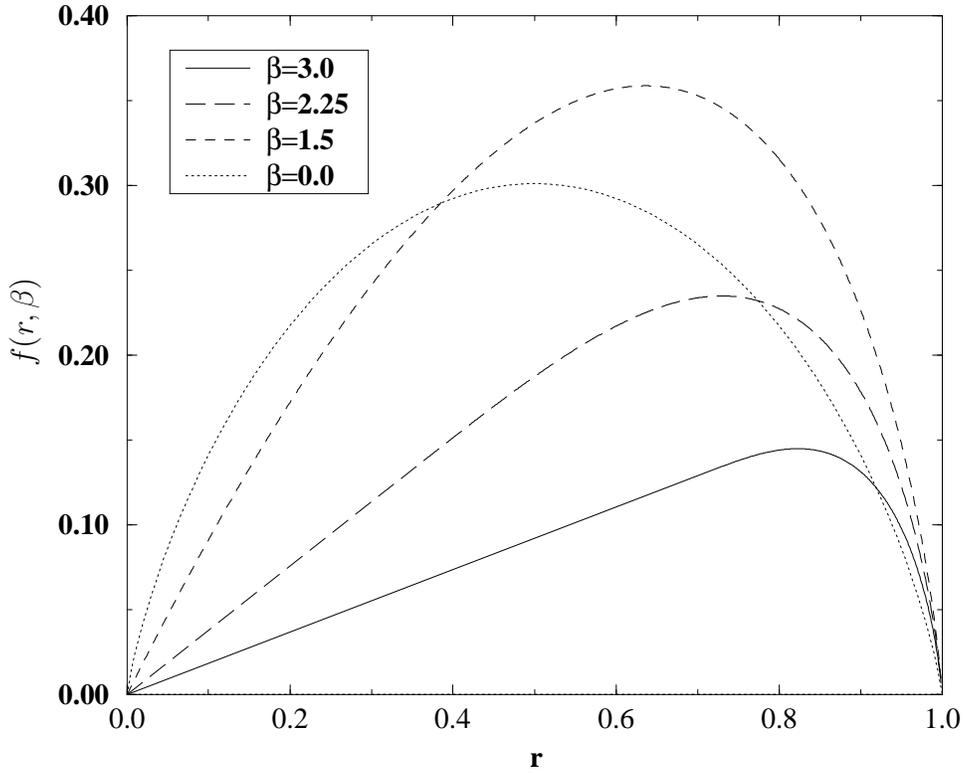,width=12cm,bbllx=51,bblly=60,bburx=516,bbury=434}
\end{center}
\caption{\label{ffr}The  $f(r,\beta)$ for various values of $\beta$. 
The curves for $\beta>2$ possess a linear part at $r=0$. The main
difference between the curves for $\beta=0$ and $\beta=1.5$ is the former 
has an infinite derivative $\partial_r f$ at $r=0$, whereas
the latter has a finite derivative. 
For all the curves the derivative at $r=1$ is infinite. See figure \ref{f1}.}
\end{figure}
The function $f(r,\beta)$ as a function of $r$ has different qualitative 
behaviour depending on $\beta$. For $\beta>2$ the function has a linear
piece for $r<r_{cr}$. This piece becomes shorter when $\beta$ approaches $2$
and finally disappears.  For $1<\beta<2$ the function has no linear piece,
but it does have a finite derivative at $r=0$. 
This distinguishes this range from
$\beta<1$ where the derivative $\partial_r f$ at $r=0$ is infinite.
Representative shapes of the function $f$ for $\beta$ from the three ranges 
are shown 
in figure~\ref{ffr}. This figure is supplemented by 
figure~\ref{f1} where the derivative is plotted to 
expose its behaviour at $r=0$.
With this picture in mind one can see the main feature of the solution
$r_*$ of the saddle point equation (\ref{trans}). Namely, for $\beta<1$, for
each $\kappa$ it has always a solution in the range $(0,1)$. For $1<\beta<2$, 
$r_*$
lies in the range $(0,1)$ so long as $\kappa>\kappa_{cr}$, whereas $r_*=0$ for
$\kappa\le \kappa_{cr}$. Finally, for $\beta>2$, the solution lies in the range
$(r_{cr},1)$ for $\kappa>\kappa_{cr}$ and then jumps to $r_*=0$ 
for $\kappa\le\kappa_{cr}$. This is illustrated in figure~\ref{f2}.
\begin{figure}
\begin{center}
\psfrag{df}{$\partial_r f(r,\beta)$}
\psfrag{r}{$r$}
\epsfig{file=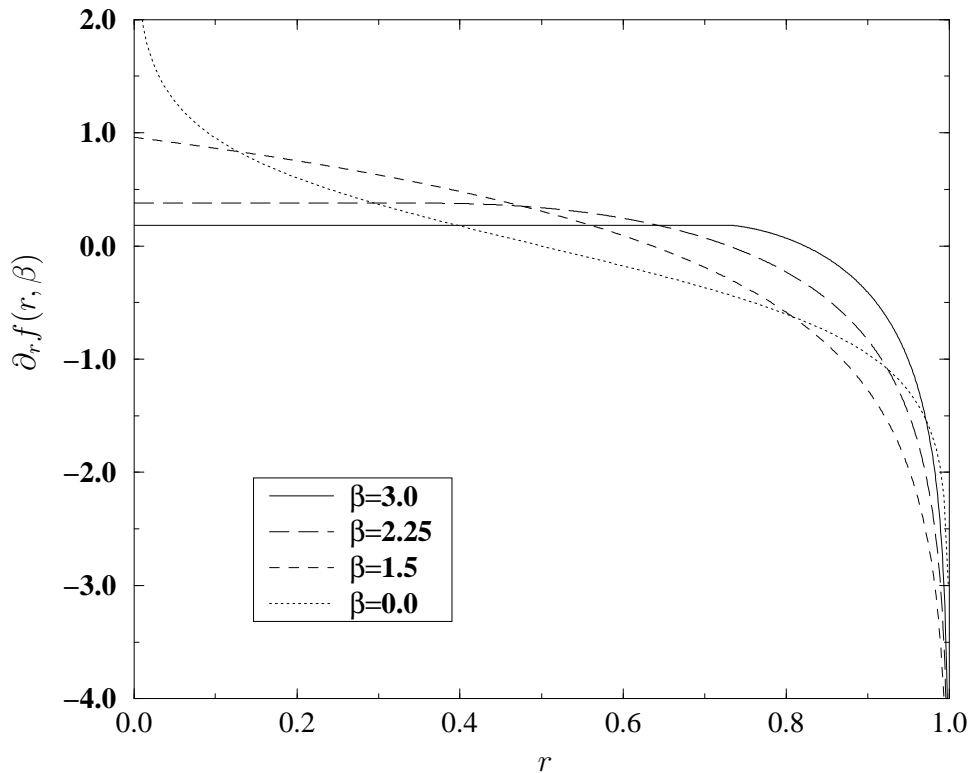,width=12cm,bbllx=51,bblly=60,bburx=516,bbury=434}
\end{center}
\caption{\label{f1} The derivative $\partial_r f(r,\beta)$ of the curves
from the figure \ref{ffr}. All go to infinity at $r=1$. Depending on the
value of $\beta$ they drastically change behaviour at $r=0$~: for $\beta=0$
they go to infinity, for all other are finite. For $\beta=2.25$ and $\beta=3.0$
they are constant in some range of $r$.}
\end{figure}
The discontinuity at $\kappa=\kappa_{cr}$
(\ref{kcr}) has the height $r_{cr}$ given by (\ref{rcr}).
The average $\langle r \rangle = r_*$ 
is an order parameter for the model. 
A simple consequence of equation (\ref{rcr})
is that the transition becomes continuous 
when $\beta=2$ since $r_{cr}=0$. In fact,
the transition stays continuous as long as
$\beta$ is in the range $(1,2]$. 
In this range the  $r_{cr}=0$ and  $\kappa_{cr}$ is finite and given by
(\ref{kcr}). 
As $\beta$ approaches unity, $\kappa_{cr} \to - \infty$ 
and the transition disappears. This process is illustrated
in figure~\ref{f2}.  
For $\beta <  1$ the values of the derivative $\partial_r f(r,\beta)$
span the interval $(-\infty,\infty)$ and the
saddle point equation (\ref{trans}) always has a  nonzero
solution $r_*$ for the whole range of $\kappa$. The
model therefore has only one phase and no phase transition.
To illustrate this, consider the solution  when $\beta=0$~:
\begin{eqnarray}
r_* = \frac{1}{1+e^{-\kappa}} \, ,
\end{eqnarray}
which is a smooth function spanning the range $(0,1)$
without any singularity.

\begin{figure}[t]
\begin{center}
\epsfig{file=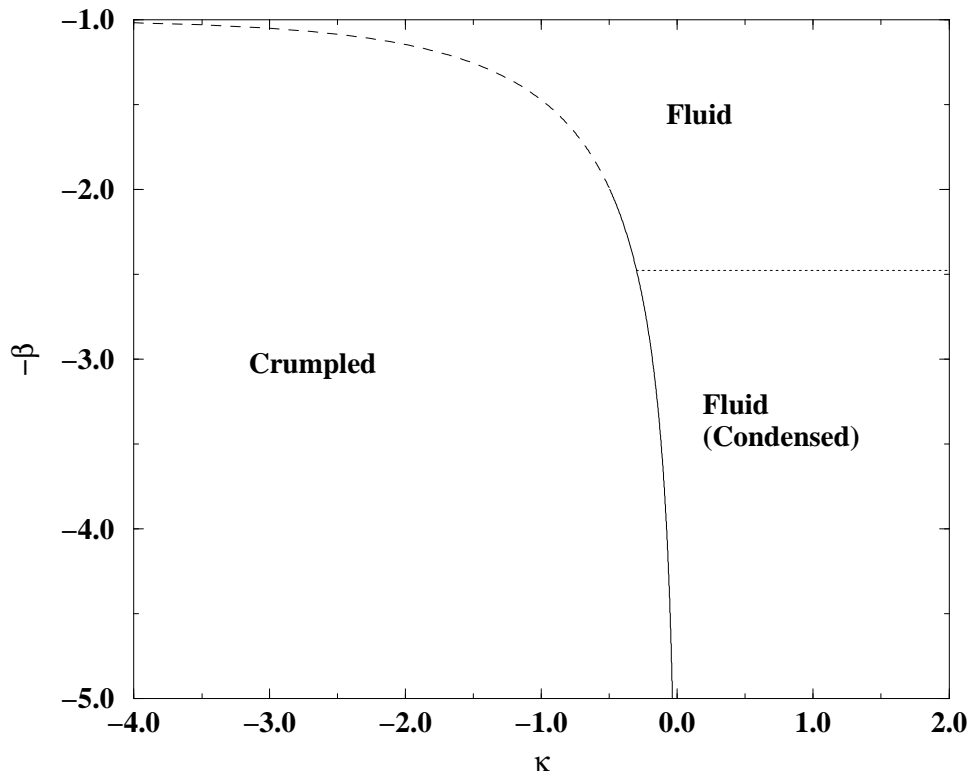,width=12cm,bbllx=51,bblly=60,bburx=516,bbury=434} 
\end{center}
\caption{\label{f3}The phase diagram. 
The continuous line denotes  a first-order phase transition. 
The dashed line denotes a continous transition. 
The dotted line shows the transition line
that appears after addition of the {\em artificial} cut-off.}
\end{figure}

The phase diagram in the $(\kappa,\beta)$ plane is
drawn in figure~\ref{f3}. 
The solid line is  a
discontinuous phase transition with a nonzero
gap $r_{cr}$ (\ref{rcr}) while the dashed line
is a continuous transition.

Coming back to the range $(1,2)$ of $\beta$ one can determine the singularity type 
of the thermodynamical functional $\phi(\kappa,\beta)$ at the transition {\em ie} 
when $\kappa \rightarrow \kappa_{cr}$. 
For small $\mu$~:
\begin{eqnarray}
g_\beta(\mu) = \Gamma(1-\beta) \mu^{\beta-1}
+\sum_{k=0}^{\infty} \zeta(\beta-k) \frac{-\mu^k}{k!} \, .
\end{eqnarray}
Inserting this into the equation (\ref{mur})
one obtains for  small $\mu$~: 
\begin{eqnarray}
r \sim \mu^{2-\beta} + \dots \quad ,
\end{eqnarray}
or, after inverting it for $\mu$~:
\begin{eqnarray}
\mu_* \sim r^{1/(2-\beta)} \, .
\end{eqnarray}
In this range of $\beta$  $r_{cr}=0$ and so $\mu_*$ is given by the
equation~:
\begin{eqnarray}
\kappa = - \log g_\beta(\mu_*)
\end{eqnarray}
For small $\mu_*$ we can 
write~:
\begin{eqnarray}
\kappa - \kappa_{cr} =- \log g_\beta(\mu_*) + \log\zeta(\beta) \sim 
\mu^{\beta-1} \sim r^{(\beta-1)/(2-\beta)} \, .
\end{eqnarray}
Inverting this with respect to $r$ we eventually get~:
\begin{eqnarray}\label{x}
\langle r \rangle = r_* \sim (\kappa - \kappa_{cr})^x, \makebox[2cm]{where}
x = \frac{2-\beta}{\beta-1} \quad .
\end{eqnarray}
The exponent $x$ varies between $\infty$ and $0$ as $\beta$ grows from one to
two. In the limiting case $\beta\rightarrow 1$, the exponent $x\rightarrow
\infty$ which means that the singularity becomes arbitrarily soft, which perfectly
matches the fact that for $\beta<1$ there is no singularity. At the other
end of the interval, when $\beta\rightarrow 2$, the exponent
$x\rightarrow0$, which means that $\langle r \rangle$ 
becomes an arbitrarily
steep function at $\kappa_{cr}$ which again agrees with the fact that for
$\beta>2$ the function is discontinuous at $\kappa_{cr}$.

What we have discussed so far is an extension 
and refinement of the analysis
of the model presented in \cite{bb}. A few comments are in order.
The analysis can be generalised for an arbitrary set of weights. 
In general, the transition takes place when the argument $e^{-\mu}$ 
of the series (\ref{kg0}) approaches the radius of convergence of the series.
Denote the radius by $e^\Delta$. In our example the radius was $1$ {\em ie} $\Delta=0$,
but in general $\Delta$ may be any finite number.
The condition for the existence of the transition is that
the generating function $K$ (\ref{kg0}) is finite at the radius of convergence 
of the series~:
\begin{eqnarray}
K(\Delta,\beta) < \infty \, .
\label{kk}
\end{eqnarray}
The family of weights may be parametrised by many
parameters. In the last formula, the parameters are represented 
by a single symbol $\beta$. The equation~:
\begin{eqnarray}
-\kappa_{cr} = K(\Delta,\beta) \, ,
\label{crl}
\end{eqnarray}
is the equation of the critical line between the phases of the
$(\kappa,\beta)$ ensemble. 
Moreover, as one can see by repeating the 
arguments previously used for the particular weights (\ref{w}), 
the condition for the transition to be discontinuous is~:
\begin{eqnarray}
\partial_\mu K(\Delta,\beta) < \infty 
\end{eqnarray}
since the inverse of the derivative $\partial_\mu K$ corresponds 
to the latent heat. The transition is continuous for $\beta$'s
for which the latent heat vanishes and condition (\ref{kk}) 
is fulfilled. 

The position of the critical line on the phase diagram can be easily changed
by a slight modification of weights.
For example, a simple multiplication by a constant
$C$~: $p(q) \rightarrow C p(q)$ corresponds to a horizontal shift of the
critical line in the phase diagram~(Fig.~\ref{f3}). In general a 
slight modification of weights can also change the $\beta$ position of 
critical line and thus the critical exponent (\ref{x}). In this respect
this model is ``non-universal'' (see also \cite{bb-bp}). 


So far we have discussed the behaviour of the model in the thermodynamic
limit. We now determine the exponents $y$ (\ref{zf}) and $Y$
(\ref{phi}). It is convenient
to define the grand canonical partition function~:
\begin{eqnarray}
{\cal Z}
(\mu,\kappa,\beta)=\sum_{N=1}^\infty e^{-N\mu} Z(N,\kappa,\beta)\sim
(\mu-\mu_0)^{-1-Y} \, ,
\end{eqnarray}
whose singularity type depends on the exponent $Y$.
The critical value of the chemical potential is
$\mu_0 = \phi(\kappa,\beta)$ is zero in the crumpled
and nonzero in the fluid phase. The value of the 
exponent $Y$ depends also on the position $(\kappa,\beta)$
in the phase diagram.
From (\ref{z}) one obtains the following expression~:
\begin{eqnarray}
{\cal Z}(\mu,\kappa,\beta)=
\frac{e^{\kappa}g_\beta(\mu)}{1-e^{\kappa}g_\beta(\mu)} \,
\label{Zfraction}
\end{eqnarray}
from which one can extract the singularity. In the fluid phase,
the singularity comes from the residual denominator and it is~:
\begin{eqnarray}
{\cal Z}(\mu,\kappa,\beta) \sim (\mu - \mu_0)^{-1} \, ,
\label{minus1}
\end{eqnarray}
from which one concludes that $Y = 0$. 

For $(\kappa,\beta)$ in the crumpled phase, the singularity 
comes from the numerator of the fraction (\ref{Zfraction})
which itself has a power like singularity when $\mu$ 
approaches $\mu_0=0$. Thus 
\begin{eqnarray}
{\cal Z}(\mu,\kappa,\beta) \sim \mu^{\beta-1}
\end{eqnarray}
and hence $Y = -\beta$. In other words, the partition
function inherits the singularity from the weights or more
precisely from the generating function $K(\mu,\beta)$.
This effect comes from the dominance of the 
singular box contribution
over the negligible entropy.  In general the singularity
need not be power like. For
example the weights~: $p(q) = \exp(-a q + b q^{1/2})$, where
$a$ and $b$ are positive constants, would lead to 
an essential singularity. In this case $N^Y$ would not be
the dominant correction to the partition function.

Let us now determine the singularity type on the critical line.
The singularity comes from the denominator in the expression
(\ref{Zfraction}). 
It changes
at $\beta=2$. For $\beta>2$ the expansion of the denominator 
in small $\mu$ begins with the linear term. Because
the numerator approaches a non-vanishing constant for $\mu\rightarrow 0$,
we have~:
\begin{eqnarray}
{\cal Z}(\mu,\kappa,\beta) \sim \mu^{-1} \, ,
\end{eqnarray}
and hence $Y=0$ as in the fluid phase. For $1< \beta < 2$ the expansion of the
denominator begins with the singular term $\mu^{\beta-1}$ (\ref{gb}),
thus
\begin{eqnarray}
{\cal Z}(\mu,\kappa,\beta) \sim \mu^{1-\beta} \, ,
\end{eqnarray}
and $Y=-2 + \beta$. 

So far we have tacitly set $q_{min}=1$, which 
meant that each box contained at least one ball. As a result, the system 
with $N$ balls could have $M$ boxes at most. Hence the maximal 
value of $r$ is $1$. If one instead chooses $q_{min}$ to be larger 
than one then the maximal $r$ is~:
\begin{eqnarray}
r_{max} = 1/q_{min} \, .
\end{eqnarray}
In this case, the analysis of the model proceeds exactly as before, except
that in place of $1$ for the upper limit one sets $r_{max}=1/q_{min}$. 
In particular, the range of integration in equation (\ref{integ}) is now
$(0,r_{max})$ and the derivative $\partial_r f$ is infinite at $r_{max}$.
The average $\langle r \rangle$ approaches asymptotically $r_{max}$ 
when $\kappa \rightarrow -\infty$. 
So the only essential difference to the previous analysis is the
range of $r$. In particular, the phase structure of the model
is unchanged. We shall call $r_{max}$ the {\em natural limit} for $r$. 

\begin{figure}[t]
\begin{center}  
\epsfig{file=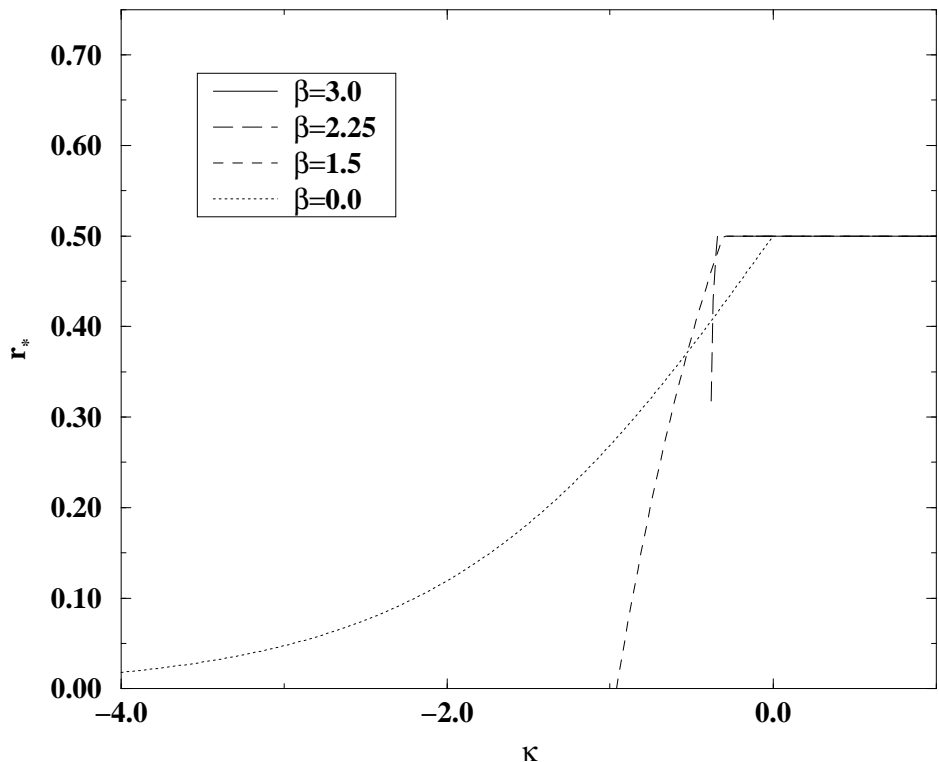,width=12cm,bbllx=51,bblly=60,bburx=516,bbury=434}
\end{center}
\caption{\label{f4}The dependence $r_*$ on $\kappa$ for various values of $\beta$
  in presence of the {\em artificial} cut-off}
\end{figure}

The phase structure of the model can be made more complex by introducing
a slight modification. Let us impose a kinematic 
upper limit on $r$. We shall call 
this the {\em artificial limit} and denote it by $r_{art}$. The 
artificial limit obviously has to lie below the natural one, $r_{art}<r_{max}$. Now
the partition function becomes~:
\begin{eqnarray}
N^{Y} e^{ N \phi(\kappa,\beta)} = 
N^{y+1} \, \int\limits_{0}^{r_{art}} \mbox{d}r e^{N (f(r,\beta) + r\kappa)} \, .
\label{integres}
\end{eqnarray}
The main difference between the natural and the artificial limit is that the
derivative $\partial_r f$ is infinite at the former and finite at the latter.
In the case of a model with the natural limit, 
this limit is
never reached by the average $\langle r \rangle$, or at worst reached only
asymptotically for $\kappa \rightarrow -\infty$, whereas the artificial limit
can be reached for finite $\kappa$ when the term $r \kappa$ dominates over
$f(r,\beta)$ in the integrand (\ref{integres}).  If we fix $\beta$ and change
$\kappa$ from large positive values towards large negative, the system is first in the
condensed phase $\langle r \rangle=0$, then at the critical value 
$\langle r \rangle$ jumps to $r_{cr}$ and later smoothly rises until it
reaches $r_{art}$ where it sticks (see figure~\ref{f4}). 
Such a  situation is very similar to the
previous standard one. 

There exists, however, the possibility of quite different behaviour. 
For $\beta > \beta_*$ 
where $\beta_*$ is defined by the equation~:
\begin{eqnarray}
r_{art}=\frac{\zeta(\beta_*)}{\zeta(\beta_* - 1 )}
\end{eqnarray}
the value of $r_{cr}$ is larger than $r_{art}$.  In this case, at the transition
from the condensed phase $\langle r \rangle$ jumps from zero directly to
$r_{art}$. It cannot jump to $r_{cr}$ because this lies above $r_{art}$ and is
beyond the allowed kinematic range (figure~\ref{f4}).
This gives a new
phase which is neither crumpled nor fluid. It has $\langle r \rangle > 0$ but
on the other hand there is a singular box in the system.  We call this phase
{\em condensed}, as it is analogous to the condensed phase of the canonical
$Z(M,N,\beta)$ ensemble.  We have marked the critical line separating the 
condensed phase from fluid one by the dotted line on the phase diagram in the
figure~\ref{f3}.

Define the quantity~: 
\begin{eqnarray} 
\langle s \rangle = \frac{\langle
 q_{sing} \rangle }{N} \, , 
\end{eqnarray} 
which is the fraction of all balls which are
in the singular box. 
The quantities $\langle r \rangle$ and $\langle s\rangle$ play the role of
the order parameters. We have~: 
\begin{eqnarray}
\begin{array}{lll}
\mbox{crumpled~:} & \langle r \rangle = 0 \, , &  \langle s \rangle =1 \, , \\
\mbox{fluid~:}     & \langle r \rangle > 0 \, , &  \langle s \rangle =0 \, , \\
\mbox{condensed~:}  & \langle r \rangle > 0 \, , &  \langle s \rangle >0 \, . \\
\end{array}
\end{eqnarray} 

The phase diagram is very similar to the one of simplicial gravity with the
matter fields or the measure term \cite{bbkptt1,bbkptt2}. If one naively applies
the mean field model to 4d simplicial gravity one would get the following
constraint for the orders of vertices~:
\begin{eqnarray} 
q_1 + q_2 + \dots + q_{N_0} = 5 N_4
\end{eqnarray} 
where $N_0$ is the total number of vertices, $q_i$ the order of
vertex $i$, and $N_4$ is the number of 4simplices. We thus 
have the correspondence
$N_0 \leftrightarrow M$, $\kappa_0 \leftrightarrow -\kappa$ and $N_4
\leftrightarrow N$. The minimal order of a vertex is $q_{min}=5$, so the natural
limit $r_{max}= 1/5$. This means that~:
\begin{eqnarray} 
\frac{\langle N_0 \rangle}{N_4} = 1 
\end{eqnarray} 
is the natural upper
limit for $\langle N_0\rangle/N_4$ from the point of view of the mean field
model. As we know this limit is never reached since the simplicial structure
imposes the limit $\langle N_0 \rangle = 1/4$ which in the language of the mean
field model corresponds to the artificial upper limit\cite{w}. The full
4d theory thus 
provides the structure which we have assumed in the discussion
of the mean--field model. Of course, the
cut off at $1/4$ is not as sharp as we have assumed. 
One expects a
smooth cut off, which means that the entropy of configurations for $\langle N_0
\rangle/N_4$ approaching $1/4$ gradually decreases in comparison with the mean
field model, so that the upper limit $\langle N_0 \rangle/N_4=1/4$ is not
reached at finite $\kappa_0$ but rather asymptotically approached for large
$\kappa_0$.

The source of the limit $\langle N_0 \rangle /N_4 = 1/4$ in simplicial
gravity is purely geometrical and results from the  fact that $q_i$
is the number of 4-simplices.  A consequence of this is that if we add a point
with $q_i=5$, for example by a barycentric subdivision, we increase the numbers
$q_j$ of neighbouring vertices\cite{w}. In other words, $q_i$'s are not independent and
in particular one cannot make all of them equal to $5$ as the natural limit would
require.

The exponent $Y$ requires some comment. In the theory of random
geometries one defines the string susceptibility exponent $\gamma$,
which is the counterpart of $Y$. 
The exponents $Y$ is shifted by a constant with respect to the standard 
definition of $\gamma$~:  $Y=\gamma-3$.
If one followed these definitions in the fluid phase
one would obtain 
$\gamma=3$ in the $(\kappa,\beta)$ ensemble
since, as we have seen, $Y=0$.
However, the mean field model neglects
all the subtle effects and  combinatorial factors coming from
topological constraints imposed
by the local geometry so one should treat this value with some caution.

Similarly, the crumpled phase of the ball in boxes model
has $Y=-\beta$, but the following argument suggests
that the corresponding value of $\gamma$ is {\em not}
that which is naively expected.
The balls in boxes model in
the particular case of the $r=1/2$ ensemble can be  mapped onto the
branched polymer model with the topology of sphere \cite{bbj,bb-bp}.
One can recognise the fluid phase of the balls in boxes model
as the generic phase of the branched polymer model. For the former
$y=-1/2$, whereas for the latter $\gamma=1/2$, from which one can conjecture
$y=\gamma-1$\footnote{We have $y$ rather than $Y$ as we are dealing
with a fixed number of boxes ensemble in the branched polymer analogy.}.
Switching back to the varying number of boxes ensemble we have $Y=y$ in the crumpled 
phase from equation (\ref{Yy2}),  
and $\gamma  = 1 - \beta$.

To calculate $\gamma$ and hence $Y,y$
in the condensed phase we will use a heuristic
argument. We know that $Y=y$ since the phase is again on the kinematic
border (\ref{Yy2}), as for the crumpled phase, but $y$ is now the exponent of the fixed $r=r_{art}$
ensemble. As before, in the particular case $r=1/2$ we can take the
value of $\gamma=2-\beta$ from the branched polymer model (\cite{bb-bp}). 
If we assume that this relation holds independently of $r$, and
that $y=\gamma-1$ as conjectured above for the crumpled phase, we obtain $Y=y=1-\beta$ which differs
by one from the crumpled phase value.

We think that the main feature of the mean--field solution is 
that the exponent $\gamma$ properly reflects an universal behaviour, 
{\em ie} it is independent on the location in the phase diagram as long
as the system is in the fluid phase. This is more important in the context
of simplicial gravity than the particular 
predictions for the exponent $\gamma$ calculated for
our choice of power law weights. Mean field theories will, in any case,
generically predict critical exponents incorrectly. 

In summary, we have seen that the mean field model of simplicial quantum
gravity can account for many of the features seen in simulations of the 
full model. With the correct choice of ensemble (a varying number of boxes)
we reproduce the first order nature of the transition for 
for $\beta\in (2,\infty)$, while for $\beta \in (1,2]$ it is continuous,
disappearing altogether at
$\beta=1$.
We have also remarked that
taking further account of the {\em local} geometrical constraints
in the model can be implemented as a lowered limit on the region of integration
in the saddle point equation, which we have denoted as the artificial upper limit. With the artificial
limit in place we found the modified phase diagram of figure~\ref{f3},
which possesses a new condensed phase.
This is similar to the crinkled phase of full simplicial gravity with 
matter fields or the (apparently equivalent) measure term \cite{bbkptt1,bbkptt2}.

\section*{Acknowledgements}
We are grateful to B.~Petersson and G.~Thorleifsson for discussion. 
This paper was partially supported by KBN grants 2P03 B04412, 2P03 B00814
and by the Polish--British Joint Research Collaboration Programme under
the project  WAR/992/137.
P.B was supported by the Alexander von Humboldt Foundation.

\end{document}